\pdfoutput=1
\documentclass[12pt]{article}
\usepackage{latexsym}

\usepackage{graphicx}

\def\tr{{\rm tr}}
\def\ket#1{\mid~\!\!\!{#1}~\!\!\rangle}
\def\bra#1{\langle~\!\!{#1}~\!\!\!\mid}

\def\Q{\rm quantum }

\def\QM{\rm quantum mechanics }
\def\qm{\rm quantum mechanics}
\def\QMl{\rm quantum-mechanical }

\def\${\enspace$}
\def\M{\rm measurement }
\def\m{\rm measurement}
\def\CY{\rm complementarity }
\def\cy{\rm complementarity}
\def\PR{\rm principle }
\def\pr{\rm principle}

\begin{document}

\noindent {\large \bf Critical
Assessment of Wave-Particle
Complementarity\\ via Derivation from
Quantum Mechanics}

\vspace{0.5cm}

\noindent {\bf Fedor Herbut}

\vspace{0.5cm}

\noindent {\bf Abstract} After
introducing sketchily Bohr's
wave-particle complementarity principle
in his own words, a derivation of an
extended form of the principle from
standard \QM is performed.
Reality-content evaluation of each step
is given. The derived theory is applied
to simple examples and the extended
entities are illustrated in a thought
experiment. Assessment of the approach
of Bohr and of this article is taken up
again with a rather negative conclusion
as far as reflecting reality is
concerned. The paper ends with
quotations of selected incisive
opinions on Bohr's dogmatic attitude
and with some comments by
the present author.\\

\vspace{0.5cm}


\vspace{0.8cm} \normalsize \rm
\section{Introduction}

\noindent Recent investigation
\cite{FHinterf}, \cite{FHScully1},
\cite{FHScully2}, has shown that the
{\bf relative-reality-of-wave-function}
point of view gives good insight in
some intricate experi-

{\footnotesize \rm \noindent
\rule[0mm]{4.62cm}{0.5mm}

\noindent F. Herbut (mail)\\
Serbian Academy of Sciences and Arts,
Knez Mihajlova 35, 11000 Belgrade,
Serbia\\
e-mail: fedorh@infosky.net and from USA
etc. fedorh@mi.sanu.ac.yu}

\noindent ments. This may come as a
surprise in view of the well-known fact
that for a very long time Bohr and
Copenhagen reigned the field. One of
the leading ideas was Bohr's
wave-particle complementarity (or
duality) principle. This investigation
is {\bf aimed} at a critical derivation
of it, or rather of its natural
extension, from \qm , and at an
appraisal from the point of view of the
reality-of-state approach. A precise
criterion will emerge distinguishing
Bohr's case from the extension.

To begin with, one should remember the
often quoted commentary of Einstein
\cite{Einstein}:
\begin{quote}
{\bf Quote EINSTEIN:} "... Bohr's
principle of complementarity, the {\it
sharp formulation} of which, moreover,
I have been unable to achieve despite
much effort which I have expended on
it." (Emphasis by F. H.)
\end{quote}

Complementarity, in contrast to
indeterminacy, seems to have remained
{\it unsettled between Einstein and
Bohr}.

There were other disquieting events
concerning complementarity. As early as
in 1951 a no lesser physicist than Max
Born expressed doubts in one of his
books \cite{Born}:
\begin{quote}
{\bf Quote BORN:} "The conceptions
"particles" and "waves" have no such
complementary character (he means
Bohr's 'mutual exclusion' of the
corresponding experiments, F. H.), as
in many cases both are needed ..."
\end{quote}

Further, there is the paradox (with
respect to Bohr's wave-particle
complementarity principle) of Ghose and
Home \cite{Home} in a real experiment.
This important work, and some other
thought experiments \cite{other1},
\cite{other2} indicate that Bohr's
intuitive complementarity principle was
too narrowly conceived. It may be that
in Bohr's time one did not think of
sophisticated experiments; the simple
ones seemed quite baffling.

It is well known that the formal
structure of \QM has had unparalleled
success in predicting probabilities of
\M results in microscopic phenomena.
Not one prediction of the former was
proved false in the latter.

In an attempt to derive a sharp form of
wave-particle duality, we will deal
with {\it experiments} in which an {\it
observable} with a purely discrete
spectrum is {\it exactly measured}. In
other words, experiments with
positive-operator-valued measures
(POVM) and inexact (or unsharp) \m s
are outside the scope of this study.

We start by a glance at Bohr's idea of
the duality in question.\\

\section{On Bohr's
Wave-Particle Complementarity
Principle}

\noindent  I'll present what I think is
the most important part of Bohr's \CY
\PR in four quotes of Bohr's own words.
In the first \cite{B1}, Bohr explains
what the problem is in the example of
the well-known Mach-Zehnder
interferometer \cite{MZ} (cf the upper
part of the Figure in subsection 9.1).

\begin{quote}
{\bf Quote BOHR1:} "The extent to which
renunciation of the {\it visualization}
of atomic phenomena is imposed upon us
by the {\it impossibility of their
subdivision} is strikingly illustrated
by the following example to which
Einstein very early called attention
and often has reverted. If a
semi-reflecting mirror is placed in the
way of a photon, having two
possibilities for its direction of
propagation, the photon may either be
recorded on one, and only one, of two
photographic plates situated at great
distances in the two directions in
question, or else we may, by replacing
the plates by mirrors, observe effects
exhibiting an interference between the
two reflected wave-trains. In any
attempt at a {\it pictorial
representation} of the behavior of the
photon we would, thus, meet with the
{\it difficulty}: to be obliged to say
on the one hand, that the photon always
chooses one of the two ways and, on the
other hand, that it behaves as if it
had passed both ways." (Emphasis by F.
H.)
\end{quote}

In the next quote \cite{B2} Bohr
expounds his answer to the
"difficulty", in terms of his famous
{\it principle of complementarity}:

\begin{quote}
{\bf Quote BOHR2:} "Information
regarding the behavior of an atomic
object obtained under definite
experimental conditions may, however,
according to a terminology often used
in atomic physics, be adequately
characterized as {\it complementary} to
any information about the same object
obtained by some {\it other}
experimental arrangement excluding the
fulfillment of the first conditions.
Although such kinds of information {\it
cannot be combined into a single
picture} by means of ordinary concepts
(i. e., by assuming that the
descriptive terms in the complementary
descriptions refer to "the same
object"), they represent indeed equally
essential aspects of any knowledge on
the object in question which can be
obtained from this domain." (Emphasis
by F. H.)
\end{quote}

Another quote \cite{B3} sums up
complementarity in a less bulky way:
\begin{quote}
{\bf Quote BOHR3:} "...evidence
obtained under different experimental
conditions {\it cannot be comprehended
within a single picture}, but must be
regarded as complementary in the sense
that only the totality of the phenomena
exhausts the possible information about
the objects."
\end{quote}

In application to the Mach-Zehnder
'difficulty' (see Quote BOHR1), the
complementarity principle treats the
which-path and the interference
versions (described in Quote BOHR1) as
two complementary experiments. Hence,
the 'visualizations' or 'pictorial
representations' in them cannot be
combined. In such a manner Bohr solves
the mentioned mind-boggling perplexity.
This is so even in Wheeler's well-known
delayed-choice version
\cite{WheelerMZ}, in which the choice
between 'which-path' or 'interference'
is made after the photon has passed the
semi-reflecting mirror. This is clear
from the immediate continuation of
quote BOHR1:
\begin{quote}
{\bf Quote BOHR4:} "It is just
arguments of this kind which recall the
{\it impossibility of subdividing
quantum phenomena} and reveal the
ambiguity in ascribing customary
physical attributes to atomic objects."
(Emphasis by F. H.)
\end{quote}

In my understanding, by 'subdivisions'
(cf also quote BOHR1) Bohr means both
spatial and temporal ones.\\

\vspace{0.5cm}

\section{Introduction into
Visualization Theory}

\noindent As well known, Bohr {\it
applied classical physics to the
behavior of the macroscopic measuring
agency} that measures some observable
\$A\$ at the final moment \$t_f\$ of
some experiment. But he did more than
that. He pushed description in terms of
classical physics also into an answer
to the question 'What happens in the
experiment?'. As we saw in Quotes BOHR2
and BOHR3, he called this
'visualization' or 'pictorial
representation'. One wonders why he did
go so far with classical description.
Perhaps, we can find an answer in the
following quote \cite{B4}.
\begin{quote}
{\bf Quote BOHR5:} "The language of
Newton and Maxwell will remain the
language of physicists {\it for all
time}" (Emphasis by F. H.)
\end{quote}

For a thorough discussion of this point
see Schlosshauer \cite{Schlossh}.

Reichenbach has pointed out that
three-valued logic is more in the
spirit of \QM than the standard
two-valued one \cite{Reich}. But we do
think in terms of two-valued logic.
Perhaps for this reason, three-valued
logic did not find much application in
\QM or its philosophy. It seems to me
that Bohr believed that the situation
is similar with physics. According to
him, apparently, we cannot help viewing
the world around us in terms of
classical physics. This is one of the
reasons, perhaps why Bohr strived to
pervade \QM with classical
physics as much as possible.\\

Some {\it prerequisites} for a theory
are now going to be presented. They are
meant to be the basis of a derived
sharp form of the complementarity
principle.

Let us consider a {\bf quantum
experiment}. It begins, at an initial
moment \$t_i,\$ with a {\bf prepared}
spatial ensemble of quantum systems, or
with a single individual system from
the, experimentally and theoretically
identical, time ensemble. Both the
ensemble and each individual system in
it are described by some density
operator \$\rho(t_i)\$ (mostly by the
special case of a state vector or wave
function - as often said). The elements
of the ensemble are assumed to be
non-interacting with each other and
with the environment. Thus, due to
dynamical isolation, one has evolution
with a unitary operator
\$U(t_f-t_i,t_i)\$ in the
Schr\"{o}dinger picture. Throughout
this paper, we shall omit the moment
when the operator is applied. It will
be understood that it is at the earlier
of the two moments defining the time
interval of evolution. Thus,
$$\rho(t_f)=U(t_f-t_i)\rho(t_i)
U(t_f-t_i)^{\dag}, \eqno{(1)}$$ where
the dagger denotes adjoining (and, in
this case, adjoint equals inverse).

As it has been stated, it is assumed
that we have {\bf exact measurement} of
an ordinary, complete or incomplete,
observable with a purely discrete
spectrum at the final moment \$t_f\$.
Let \$A\$ be the Hermitian operator
representing it in the formalism, and
let its spectral form be
$$A=\sum_na_nP_n,\quad n\not=
n'\enskip\Rightarrow\enskip a_n\not=
a_{n'}.\eqno{(2a)}$$ (The index or
quantum number \$n\$  enumerates the
distinct eigenvalues \$a_n\$ and
eigen-projectors \$P_n.\$ ) The
accompanying spectral decomposition of
the identity is
$$\sum_nP_n=I,\eqno{(2b)}$$ where \$I\$
denotes the identity operator. Relation
(2b) is also called the relation of
completeness and also the closure
property. The sum in (2b) has at most a
countable infinity of orthogonal
projector terms.

In many experiments, the completeness
relation (2b) makes the spectral form
(2a) superfluous, i. e., one talks in
terms of eigen-events \$P_n\$ without
eigenvalues \$a_n$.

As it is well known, the measuring
apparatus detects a result \$a_n\$ (cf
(2a)) or, more generally put, the
occurrence of an eigen-event \$P_n\$,
in the state \$\rho(t_f)\$ for each
individual system.  As to the ensemble,
the observed relative frequency of the
result is close to the predicted
probability value, given by the
so-called trace formula (the up-to-date
equivalent of Born's rule)
$$p_n=\tr\Big(P_n\rho(t_f)\Big).
\eqno{(3)}$$ (One should obtain
equality in the limes of an imagined
infinitely large ensemble.)\\

We give now a formal definition of the
{\bf retrospective observables}
\$A^r(t)\$, \$t_i\leq t <t_f\$, the
Hermitian-operator representatives of
which have the spectral forms:
$$A^r(t)\equiv\sum_na_nP_n^r(t),
\quad n\not= n'
\enskip\Rightarrow\enskip a_n\not=
a_{n'},\eqno{(4a)}$$ with the
definition of {\bf retrospectivity} at
the moment \$t\$:  $$\forall n:\quad
P_n^r(t)\equiv
U^{\dag}(t_f-t)P_nU(t_f-t).
\eqno{(4b)}$$ (It may be sometimes
suitable to allow the eigenvalues of
\$A^r(t),\$ if they have a physical
meaning, not to be necessarily equal to
those of \$A\$ because it is the
projectors that count.) Naturally, (2b)
implies the completeness relation
$$\sum_nP_n^r(t)=I.\eqno{(4c)}$$\\

Besides the actually measured
observable \$A\$ and its formal images
\$A^r(t)\$, which are 'back-evolved' to
a moment \$t\$, let also new
observables \$B(t)\$, 'jokers' for the
time being, be introduced. They will
play a natural role in completing the
forthcoming derivation.

Let also the observables \$B(t)\$ be
defined in spectral form
$$B(t)=\sum_kb_kQ_k(t),\quad k\not=
k'\enskip \Rightarrow\enskip b_k\not=
b_{k'}, \eqno{(5a)}$$ with the
completeness relation
$$\sum_kQ_k(t)=I\eqno{(5b)}$$ for the
eigen-projectors. We consider \$B(t)\$
as given at the initial or some
intermediate moment \$t$.

It will also be useful to formally
evolve \$B\$ to the final moment
\$t_f\$ when the actual \M of \$A\$
takes place. One has
$$B^f\equiv U(t_f-t)B(t)
U(t_f-t)^{\dag},\eqno{(6a)}$$
$$\forall k:\quad
Q_k^f=U(t_f-t)Q_k(t)
U(t_f-t)^{\dag}\eqno{(6b)}$$ (cf the
convention adopted above (1)).

One should remember that it is the
Schr\"{o}dinger picture that is being
made use of. The evolution (or 'back
evolution') of an observable are
auxiliary concepts (nothing to do with
the Heisenberg dynamical picture).\\

\section{Blindness to
Coherence in Measurement}

\noindent Let us return to the actually
measured observable \$A\$ in the final
state \$\rho(t_f)\$, and let us define
the {\bf coherence-deprived mixture}
$$\rho(t_f)_M\equiv\sum_np_n\rho(t_f)_M^n
\eqno{(7a)}$$ {\bf corresponding to}
\$A\$, where the (statistical) weights
are the probabilities \$p_n\$ given by
(3), and
$$\forall n, p_n>0:\quad
\rho(t_f)_M^n\equiv
P_n\rho(t_f)P_n\Big/p_n\eqno{(7b)}$$
are the constituent states of which the
mixture consists (concerning \$n\$, cf
(2a)). One should note that if
\$p_n=0,\$ then, as it is usually
understood, the entire corresponding
term in (7a) is zero - in spite of the
fact that the corresponding density
operator (7b) is not defined. (This
will not be mentioned again for other
formal mixtures below.)

In each state defined by (7b) the
observable \$A\$ has the definite
corresponding value \$a_n\$. This is so
because \$\tr\Big(P_n\rho(t_f)_M^n
\Big)=\tr\Big(\rho(t_f)_M^n\Big)=1\$.

It is possible {\bf coherence} that
makes a difference between a given
state \$\rho(t_f)\$ and the
corresponding mixture (7a). It is
important to be aware that 'coherence'
is a {\bf relative concept} that
applies to a state in relation to an
observable (for more details see
\cite{FHcoh}). In this case we have
possible coherence in \$\rho(t_f)\$ in
relation to \$A\$.

In view of the fact that one can always
write \$\rho = \sum_k\sum_{k'}Q_k\rho
Q_{k'}\$ (cf the completeness relation
(5b)), one can, in general, define
coherence as follows.\\

\noindent {\bf Definition 1.} A state
\$\rho\$ is coherent with respect to an
observable \$B\$ if \$\sum_{k\not=
k'}Q_k\rho Q_{k'} \not= 0\$ (cf (5a)),
or, equivalently, if \$\exists k\not=
k':\enskip Q_k\rho Q_{k'}\not=
0$.\\

In (7a) the state is by 'brute force'
deprived of all possible coherence
among the distinct values of \$A\$ in
\$\rho(t_f)=\sum_{n,n'}
P_n\rho(t_f)P_{n'}\$. (As to a measure
for the amount of coherence, see
\cite{FHcoh}.) This is why Bell calls
(7a) the 'butchered' version of
\$\rho(t_f)\$ \cite{Bell1}.

It is a {\bf crucial fact} that, in
spite of the 'butchering', one obtains
{\bf the same individual-system
results} and {\bf the same relative
frequencies} when \$A\$ is measured in
\$\rho(t_f)\$ or in \$\rho(t_f)_M\$ (cf
(7a)). Thus, the measuring apparatus
{\bf cannot distinguish} the final
state \$\rho(t_f)\$ from the
corresponding 'butchered' mixture (7a)
in the given experiment
, i. e., it is
{\bf blind to the possible coherence}
in \$\rho(t_f)$.\\

Next, let us define the mixture
corresponding to the initial or
intermediate state \$\rho(t)\equiv
U(t-t_i)\rho(t_i)U(t-t_i)^{\dag}\$,
\$t_i\leq t<t_f\$, and the
corresponding retrospective observable
\$A^r(t)\$ (cf (4a) and (4b)). But
first let us establish (by inserting
the evolution operator) that the
initial or intermediate and the final
probabilities are equal:
$$\forall n:\quad \tr\Big(P_n^r(t)
\rho(t)\Big)=\tr\Big[\Big(
U(t_f-t)^{\dag}P_nU(t_f-t)\Big)\Big(
U(t_f-t)^{\dag}\rho(t_f)U(t_f-t)\Big)
\Big]=$$ $$\tr\Big(P_n
\rho(t_f)\Big)=p_n$$ (cf (1), (3) and
(6a)).

The 'butchered' mixture that we want to
write down is:
$$\rho(t)_M\equiv
\sum_np_n\rho(t)_M^n,\eqno{(8a)}$$
where $$\forall n,p_n>0:\quad
\rho(t)_M^n \equiv
P_n^r(t)\rho(t)P_n^r(t)\Big/p_n
\eqno{(8b)}$$ (cf (4b)).

The unitary evolution by \$U(t_f-t)\$
takes the initial or intermediate
mixture (8a) into the final mixture
(7a). Therefore, in the given
experiment the state \$\rho(t)\$ and
the corresponding 'butchered' mixture
\$\rho(t)_M\$ in relation to
\$A^r(t),\$ {\it cannot be
distinguished} neither on the
individual-system level, nor as
ensembles.\\

\section{The 'Simplest
Which-Result' Visualization Theory}

\noindent The retrospective observable
\$A^r(t)\$ (cf (4a)) is, in general,
just a mathematical construction. But
in some experiments it has a {\it
physical meaning}. Then we have the
'simplest which-result' visualization.
It is slightly more general than {\it
the case of Bohr's particle-like
visualization} (see below).\\

The {\bf visualization} at issue
consists of {\bf two drastic imagined
steps of changes}.

{\bf (i)} Whereas the true initial or
intermediate state \$\rho(t),\$ which,
in general, contains {\it coherence} in
relation to \$A^r(t)\$, describes both
a laboratory ensemble and each
individual system in it, in the
visualization it is {\bf the
'butchered' mixture} \$\rho(t)_M\$ (cf
(8a)) without the possible coherence
that {\bf describes the ensemble}.

{\bf (ii)} Resorting to the so-called
'ignorance-interpretation' of a mixture
used in classical physics, {\bf the
individual system is described by one
of the constituent states}
\$\rho(t)_M^n\$ in the mixture (cf
(8b)).

In visualization, the ensemble state
\$\rho(t)_M\$, given by (8a), is a
mixture of the individual-system states
states given by (8b).

It is important to keep in mind that
the state (8b) {\it has the definite
property} \$a_n\$ of the retrospective
observable \$A^r(t)\$ (cf (4a-c)), or,
equivalently, that the eigen-event
\$P_n^r(t)\$ of \$A^r(t)\$ {\it occurs}
in it.

The 'simplest which-result' {\it
visualization}, which is being
expounded, {\it is based on the
following relations}, which are easily
seen to follow from the definitions
(4a) and (4b):
$$A^r(t)\equiv U(t-t_i)
A^r(t_i)U(t-t_i)^{\dag}=\sum_na_n
U(t-t_i)P_n^r(t_i)U(t-t_i)^{\dag}=
\sum_na_nP^r(t)\eqno{(9a)}$$ (cf (4a)
and (4b)), and
$$\forall n,\enskip p_n>0:
\quad\rho(t)_M^n=U(t-t_i)\rho(t_i)_M^n
U(t-t_i)^{\dag}=$$ $$U(t-t_i)\Big(
P_n^r(t_i)\rho(t_i)P_n^r(t_i)\Big/p_n
\Big) U(t-t_i)^{\dag}=$$ $$
\Big(U(t-t_i)P_n^r(t_i)
U(t-t_i)^{\dag}\Big)
\Big(U(t-t_i)\rho(t_i)U(t-t_i)^{\dag}
\Big/p_n\Big) \Big(U(t-t_i)P_n^r(t_i)
U(t-t_i)^{\dag}\Big)=$$
$$P_n^r(t)\rho(t)P_n^r(t)\Big/p_n,
\eqno{(9b)}$$ where \$t_i\leq t\leq
t_f\$, and \$A^r(t_f)=A\$. Thus, the
individual system is at each instant
\$t\$ imagined to be in a state
\$\rho(t)_M^n\$, which has the definite
value \$a_n\$ of \$A^r(t)\$.

In most cases \$A\$ and all its
'back-evolved' forms \$A^r(t)\$ are
localization observables. Then, one
speaks of 'which way' instead of 'which
result', and one imagines that the
system behaves as a particle moving
along a trajectory (particle-like
behavior). If, in addition, the initial
observable \$A^r(t_i)\$ has the
physical meaning of localization, i.
e., if the \QMl 'trajectory' begins at
the initial moment, then one has {\bf
Bohr's particle-like behavior} in the
famous wave-particle duality relevant
for a large number of experiments.\\

\vspace{0.5cm}

\section{A More Practical Criterion}

\noindent {\bf Remark 1} It is easy to
see that in the 'simplest which-result'
case that we consider, \$A^r(t_i)\$
satisfies the assumptions of
premeasurement (\cite{BLM} and
\cite{FHmeas}): \$A^r(t_i)\$ appears to
be the measured observable, \$A\$ the
'pointer observable', and the
macroscopic (classically described)
measuring agency appears to play the
role of objectification (or 'reading'
the result).\\

It is not always easy to evaluate the
action of the formal back-evolving
operator \$U(t_f-t)^{\dag}\$ on the
measured observable \$A\$. Hence a more
practical criterion is desirable.

We resort to the 'joker' observable
\$B\$ (cf (5a)), and make use of it
having in mind that 'simplest
which-result' visualization consists in
the fact that there exists an
observable \$B\$ with physical meaning
and the equality
\$B=A^r(t)\$ holds true.\\

\noindent {\bf Theorem 1} {\it A
necessary and sufficient {\bf
condition} for a 'which-result'
visualization}: If an observable
\$B(t)\$, \$t_i\leq t<t_f\$, has
physical meaning and, having in mind
the quantum numbers of \$B(t)\$ and
\$A\$ (cf (5a) and (2a) respectively),
there exists a bijection (a one-to-one
'onto' map) \$k\rightarrow n=n(k)\$
such that for each value of the index
\$k,\$ whenever a state \$\rho(t)\$ has
the property \$b_k,\$ or, equivalently,
the event \$Q_k\$ occurs in
\$\rho(t),\$ then the final state
\$\rho(t_f)\$ gives the result
\$a_{n(k)}\$ in the \M of \$A\$ {\bf
with certainty}, i. e.,
$$\tr\Big(\rho(t)Q_k(t)\Big)=1\quad
\Rightarrow\quad\tr\Big(\rho(t_f)
P_{n(k)}\Big)=1.\eqno{(10)}$$

If the condition is valid, then
\$B(t)=A^r(t)\$, and one has
'which-result' visualization.\\

Proof is given in Appendix A.\\

\noindent {\bf Remark 2} Theorem 1
implies that the retrospective
observable \$A^r(t)\$, and no other
observable (up to the eigenvalues
\$a_n\$, which are irrelevant), has the
required property. The property itself
can be experimentally demonstrated (see
the examples below) by showing that any
initial state with any definite value
\$b_k\$ of a given observable \$B\$
with a physical meaning necessarily
ends up in a state with the definite
value \$a_{n(k)}\$ of the actually
measured observable \$A.\$ If the
'which-result' experiment is understood
as measurement of \$A^r(t_i)\$ via the
'pointer observable' \$A\$ (cf Remark
1), then the theorem says that the
latter measures precisely one
observable (up to arbitrary eigenvalues).\\

Incidentally, I have discussed backward
'projection' in time of an actually
measured observable in two previous
articles in contexts different from the
present one \cite{FHmeas}, \cite{FHback}.\\

\vspace{0.5cm}

\section{Completion of the
'Interference' or 'Which-Result'
Visualization}

\noindent {\bf Remark 3} To develop a
full visualization theory, one needs to
answer {\bf two questions}:

{\bf (i)} Can one have a 'which-result'
experiment without one of the
retrospective observables \$A^r(t)\$
being physically meaningful?

{\bf (ii)} If one does not have a
'which-result' experiment, does one
{\it ipso facto} have a wave-like or
'interference' pictorial
representation?\\

To answer the two questions in Remark
3, we turn to the other alternative in
Bohr's wave-particle complementarity
\pr : to the {\bf wave}.

Classical waves are essentially
different from {\it quantum-mechanical
wave-like behavior}, which is universal
and contained in the very time
evolution (1) of any \QMl state
\$\rho(t_i).\$ Let us take as a simple
example diffraction of a photon through
one hole. (This case puzzled Einstein
in Solvey 1927 \cite{Solvey}.)

On passing the hole, the evolution of
the photon can be understood in a
simplified, coarse-grained way as
consisting of a diverging coherent
bundle of component probability
amplitudes constituting a half-sphere.
When the photon is located, one of the
components is realized, and all the
others just disappear, become
mysteriously extinguished. (This is the
collapse version of \QMl insight.)

Locating a photon in a dot out of a
half-sphere is only quantitatively (not
qualitatively) different from the case
of the Mach-Zehnder which-way device.
(In the latter only one component is
extinguished.)

Returning to {\it classical wave-like
behavior}, here we have components that
are {\it real} in the classical sense.
Their reality is detectable: all
components simultaneously influence the
result of measurement (or
measurements). This idea leads us to
the basic Definition 3 below, which
distinguishes 'which-result'
experiments from 'interference'
ones.\\

Coherence (see Definition 1) is usually
detected as interference. Let \$B\$ (cf
(5a)) be an observable that has
physical meaning at some moment
\$t,\enskip t_i\leq t\leq t_f\$ in the
experiment.\\

\noindent {\bf Definition 2} A state
\$\rho(t)\$ exhibits {\bf interference}
in relation to an observable \$B\$ in
the measurement of an observable \$A\$
(cf (2a)) if for at least one
eigenvalue \$a_n\$ of \$A\$ the state
\$\rho(t)\$ predicts a {\it different
probability} than {\bf the 'butchered'
mixture corresponding to} \$B\$:
$$\rho(t)_{M,B}\equiv
\sum_kp_k\rho(t)_{M,B}^k\enskip,
\eqno{(11a)}$$ where the statistical
weights are
$$\forall k:\quad p_k\equiv\tr(Q_k
\rho(t))\eqno{(11b)}$$ (cf (5a)), and
the definite-result states of \$B\$ are
$$\forall k,p_k>0:\quad
\rho(t)_{M,B}^k\equiv Q_k\rho(t)
Q_k\Big/p_k.\eqno{(11c)}$$\\

In Definition 2 "predicts" is short for
"the corresponding time-evolved state
\$\rho(t_f)\$ predicts". Thus, on the
one hand we have
\$\tr\Big(\rho(t_f)P_n\Big)\$ (cf (2a))
with \$\rho(t_f)=U(t_f-t))\rho(t)
U(t_f-t))^{\dag}\$. On the other hand,
we have
\$\tr\Big[\Big(U(t_f-t))\rho(t)_{M,B}
U(t_f-t))^{\dag}\Big)P_n\Big]\$.
Interference has set in if these two
probabilities differ for at least one
value of \$n$.\\

\noindent {\bf Definition 3 A)} We say
that an observable \$B\$ (cf (5a)) with
a physical meaning at some moment
\$t,\enskip t_i\leq t\leq t_f\$, is a
{\bf 'which-result' observable} in the
given experiment that ends in the
measurement of \$A,\$ and the
experiment is a 'which-result' one {\bf
in relation to} \$B\$, if the final
state \$\rho(t_f)\$ exhibits no
interference in comparison with the
corresponding mixture \$U(t_f-t)
\rho(t)_{M,B}U(t_f-t)^{\dag}\$ (cf
Definition 2)).

{\bf B)} We say that an {\bf
observable} \$B\$ with a physical
meaning at a moment \$t\$ is an {\bf
'interference' one}, and the {\it
experiment} is an {\it 'interference'
one} {\bf in relation to \$B\$} if
there exists at least one physically
meaningful initial state \$\rho(t_i)\$
that shows interference with respect to
\$B\$ in the measurement of \$A\$.
(Needless to say that this
\$\rho(t_i)\$ then must {\it a
fortiori} contain coherence with
respect to \$B\$.)\\

The first question in Remark 3 is now
going to be answered by deriving a
necessary and sufficient condition for
'which-result' visualization. The
condition will simultaneously answer
also the second question in the
affirmative.\\

\noindent {\bf Theorem 2} {\it
'Interference' or 'which-result'
visualization} Let \$B(t)\$ be a
physically meaningful observable at
some moment \$t,\enskip t_i\leq t<
t_f\$ in the given experiment.

{\bf A)} The observable \$B(t)\$ is a
{\bf 'which-result'} one and the
experiment is of the same kind in
relation to \$B\$ {\bf if and only if}
its evolved form \$B^f\equiv B(t_f)\$
(cf (6a)) and \$A\$ are compatible, i.
e., they {\bf commute} as operators
$$\Big[B^f,A\Big]=0. \eqno{(12)}$$

In more detail, any initial state
\$\rho(t_i)\$ and the corresponding
'butchered' state \$\rho(t_i)_{M,B}\$
(cf (11a-c)) predict the same
probability for every value \$a_n\$ of
the actually measured observable \$A\$
if and only if (12) is valid.

{\bf B)} If \$B(t)\$ is not a
'which-result' observable, then {\it
ipso facto} \$B(t)\$ and the experiment
in relation to it are {\bf
'interference'} ones. More precisely,
in this case there exists, in
principle, a physically meaningful
initial state \$\rho(t_i)\$ such that
it 'predicts' at least for one result
\$P_n\$ of \$A\$ a different
probability than the corresponding
'butchered' mixture \$\rho(t)_{M,B}\$
.\\

Proof is given in Appendix B.\\

\noindent {\bf Remark 4} One should
note the {\bf relative character} of a
'which-result' or 'interference'
experiment. But one can say, in the
spirit of Bohr's wave-particle
complementarity principle, if there
exists at least one physically
meaningful observable \$B\$ at some
instant \$t\$, \$t_i\leq t\leq t_f\$ in
relation to which the experiment is a
'which-result one', then the experiment
can be viewed as such in an absolute
sense.\\

\noindent {\bf Remark 5} A physically
meaningful observable \$B\$ that is not
the back-evolved measured observable
\$A\$ is most useful when to each
eigenvalue \$a_n\$ of \$A\$ corresponds
one eigenvalue \$b_k\$ of \$B\$ (more
precisely, each range of \$P_n\$, cf
(2a), is part of one range of some
\$Q_k^f\$, cf (12)). But one and the
same \$b_k\$ should corresponds to more
than one \$a_n\$, or else \$B\$ is as
good as \$A\$ itself (then the \$Q_k\$
and \$P_n\$ coincide), and \$B\$ is
actually the back-evolved \$A$.\\

\noindent {\bf Remark 6} If a
physically meaningful observable \$B\$
has more than two eigenvalues, then it
has nontrivial functions as new
observables. Then, it may happen that
the experiment has a different
character ('interference' or
'which-result' one) for \$B\$ and for
one of its functions.\\

\noindent {\bf Remark 7} One should
notice that, by definition, we have the
'interference' alternative if there
exists at least one initial state
\$\rho(t_i),\$ possessing coherence
with respect to the considered
physically meaningful observable \$B,\$
that is detectable as interference (see
Definition 3). If \$B\$ is an
'interference' observable, there still
may be initial states for which the
'which-result' visualization is
applicable.

One can see this, e. g., in some
quantum erasure experiments. See the
beautiful real (random delayed-choice)
experiment in \cite{origScully2}. The
photon entering the Young two-slit
experiment (the 'second' one) has
another photon (the 'first' one)
correlated to it moving in the opposite
direction. By suitable measurements on
the latter, the ensemble of all
'second' photons is broken up into two
subensembles (improper mixture of two
states), one giving an 'interference'
experiment,
and the other being a 'which-way' one.\\

\vspace{0.5cm}

\section{Simple Examples of
Visualizations}

\noindent The visualization theory
presented in sections 3-7 is now going
to be illustrated on four simple and
well-known examples, all belonging to
the {\it binary case}, i. e., to the
case when the measured observable \$A\$
has only two values.\\

\subsection{Mach-Zehnder}

\noindent Imagining the propagation of
the photon through the Mach-Zehnder
interfering device \cite{MZ} (cf the
upper part of the Figure in subsection
9.1), it traverses the first and the
second beam splitter ('semi-reflecting
mirrors' in quote BOHR1).

To understand the {\bf two
complementary experiments} to be
described, one should have in mind that
the first beam splitter can be in
place, can be removed, and can be
replaced by a totally-reflecting mirror
in the same position. When it is in
place, besides being at the standard
angle \$45^0,\$ it can be at any angle
\$0\leq\theta\leq 180^0\$. Thus, it
plays the role of a preparator. The
photon leaves the preparator at the
initial instant \$t_i$.

If the second beam splitter is removed,
we have the {\bf Mach-Zehnder which-way
device}, and in it one of the two
experiments, which we call the {\it
which-way one} - a special case of a
which-result experiment. If the second
beam splitter is in place, we have the
{\bf Mach-Zehnder interference device},
and in it the complementary experiment,
which we call {\it the interference
experiment}. At the final instant
\$t_f,\$ the photon leaves the place of
the second beam splitter (or the beam
splitter itself if it is in place) to
enter one of the detectors. Detection
at one or the other of the detectors
means occurrence of the eigen-events
\$P_1\$ or \$P_2\$ of the measured
localization observable \$A\$ (cf
(2a)). (The eigenvalues \$a_n\$ are
arbitrary and irrelevant. The results
are, this time, expressed in terms of
the localization eigen-events
\$P_i,\enskip i=1,2.\$)

In the {\bf which-way experiment} the
eigen-events \$Q_1\equiv Q_h\$
(horizontal propagation) and
\$Q_2\equiv Q_v\$ (vertical
propagation) of the physically
meaningful observable \$B\$ (cf (5a)),
which occur in the preparator, this
time coincide with those of the
retrospective observable \$A^r(t_i)\$
(cf (4a)) {\it mutatis mutandis}.

Namely, the event \$Q_h\$ takes place
if the first beam splitter is removed
and the photon propagates {\it
horizontally}. The event \$Q_v\$ occurs
if the first beam splitter is replaced
by an equally positioned mirror. Then
the photon is reflected and it
propagates {\it vertically}. Since, by
definition of the experiment, the
second beam splitter is removed, it is
obvious that the condition of Theorem 1
is satisfied.

Therefore, if the first beam splitter
is in place at some mentioned angle,
and we have a {\bf coherent initial
state}
$$\ket{\phi,t_i}\equiv \alpha
\ket{\phi}_h+\beta\ket{\phi}_v,
\eqno{(13a)}$$ where \$\ket{\phi}_h\$
is a pure state propagating
horizontally, and \$\ket{\phi}_v\$ is
one propagating vertically, and
$$|\alpha|^2+|\beta|^2=1,\quad \alpha
\not= 0\not=\beta,\eqno{(13b)}$$ then
the experiment cannot distinguish it
from the (incoherent) mixture
$$|\alpha|^2\ket{\phi}_h\bra{\phi}_h+
|\beta|^2\ket{\phi}_v
\bra{\phi}_v.\eqno{(14)}$$ This implies
the {\bf which-way visualization}.

Simply put: in spite of the first beam
splitter being in place (under some
angle), and coherence existing in the
initial state, the photon appears to
have left the first beam splitter
either horizontally or vertically (not
both at the same time). This is a
'pictorial representation' (to use
Bohr's term, cf quote BOHR1) along
classical space-time lines.

We are dealing here perhaps with the
most simple case of Bohr's
particle-like behavior.

Incidentally, one sometimes uses the
expression "the photon has which-path
information". I think, this is
thoroughly misleading because it
suggests that 'going one path' for a
single photon is a real event in
nature. But it isn't.\\

In the {\bf 'interference' experiment}
the second beam splitter is in place.
The actually measured observable \$A\$
and the physically meaningful
observable \$B,\$ or rather its
eigen-events \$Q_h,Q_v\$ (in the
preparator) are defined as in the
described complementary 'which-way'
experiment. But this time, as easily
seen, the condition of Theorem 1 is not
satisfied, and the retrospective
observable \$A^r(t_i)\$ is not equal to
B. The former observable has no
physical meaning.

Resorting to Theorem 2, it is not easy
to see if \$B^f\$ and \$A\$ commute or
not. It is easier to utilize the very
definition of 'interference'
experiments (cf Definition 2). Thus, it
is obvious that \$Q_h\$ and \$Q_v\$
contribute coherently to the two
detection rates and one has
interference.

In this case there is no visualization
or classical space-time picture in
terms of a one-way motion. One does
speak, instead, of the photon taking
both paths simultaneously in spite of
the coherent initial state (13a), but
this is only putting in words the \QMl
evolution (cf (1)).\\

It is important and satisfying to know
that both {\it single-photon}
Mach-Zehnder experiments discussed are
no longer in the realm of thought
experiments; they have become real
experiments performed in a convincing
way in the laboratory
\cite{realMZ}.\\

\subsection{Two Slits}

\noindent To apply the visualization
theory from sections 3-7 to this case,
we take the more sophisticated
Wheeler's delayed-choice version
\cite{Wheeler2slits}. The photon that
has passed the slits goes through
lenses that make the separate one-slit
paths cross at, what we call, the
'close distance', and afterwards
diverge, so that at a 'farther
distance' there is no possibility of
interference. If one puts detectors at
suitable places there, at the 'farther
distance', they detect precisely the
photon from one or the other of the
slits. We add to this independently
movable shutters on the slits for our
purposes. Thus, the {\bf which-way
experiment} is defined.

The measured observable \$A\$ is the
detection of localization at the
'farther distance'. The physically
meaningful observable \$B\$ is, as
easily seen, 'going through the one or
through the other slit'. The condition
of Theorem 1 is, obviously, satisfied,
\$B=A^r(t_i)\$, and we have {\bf
which-way visualization}, in
particular, Bohr's particle-like
behavior.

In the {\bf interference experiment}
the photons never reach the detectors
from the preceding complementary
experiment because a second screen with
detectors (or a film plate) is raised
at the 'close distance', where
interference takes place. The actually
measured observable \$A\$ is again
localization, and the physically
meaningful observable \$B\$ is the same
as in the above complementary
experiment. The retrospective
observable \$A^r(t_i)\$ is now a
complicated mathematical construction
devoid of physical meaning because the
condition in Theorem 1 is not
satisfied. Hence, there is no
'which-way' visualization.

One speaks of the photon going through
both slits, but this is only putting in
words what the evolution operator in
the \QMl formalism does.\\

\subsection{Stern-Gerlach}

\noindent In the Stern Gerlach
spin-projection measurement of a
spin-one-half particle, complementarity
comes from different axis orientations.
But for any given orientation, the
experiment {\it allows visualization}.
The measured observable \$A\$ is
defined by the dots on the screen. (It
is again a localization measurement as
in the preceding cases.) The
retrospective observable \$A^r(t_i)\$
is determined by definite spin-up and
definite spin-down before entering the
magnetic field. As easily seen, the
condition in Theorem 1 is satisfied and
we have 'which-result' visualization
though it is not of a space-time
nature. It consists in saying that the
particle has a definite spin-projection
(up or down), not both, throughout the
experiment.\\

\subsection{Double Stern-Gerlach}

\noindent Let us imagine a modified
Stern-Gerlach device measuring the
\$z$-projection of a spin-one-half
particle, but without the screen (on
which the dots should appear). Instead,
the particle just leaves in the upper
or in the lower half-space entering one
of two suitably placed second
Stern-Gerlach devices the upper one
measuring the \$x$-projection of spin,
and the lower one the \$y$-projection
(both supplied with screens giving the
dots).

It is intuitively obvious that,
whatever the coherent state entering
the first (modified) Stern-Gerlach
device, if one obtains, e. g., an upper
dot in the upper second Stern-Gerlach
device, the particle must have passed
through the upper half-space in the
first Stern-Gerlach (otherwise it would
not have reached the upper second
Stern-Gerlach). Naturally, an analogous
argument holds true for any other dot
in the second upper or lower
Stern-Gerlach device.

But this so obvious classical reasoning
is precisely an example of non-Bohrian
'which-way' visualization. Passing the
upper or lower half-space in the first
Stern-Gerlach modified device defines
the eigen-events \$Q_1\$ and \$Q_2\$ of
a physically meaningful observable
\$B\$ (cf (5a)) respectively, which
{\bf does not equal} a back-evolved
form of \$A\$. Namely, it is easy to
see that the condition in Theorem 1 is
not satisfied: Passing the mentioned
upper half-space does not guarantee
that the particle will end up in an
upper dot in the upper second
Stern-Gerlach, etc. But \$[B^f,A]=0,\$
the condition in Theorem 2 is, clearly
satisfied (spatial degrees of freedom
of a particle and its spin ones always
commute). Thus, we have here an example
of a 'which-way' observable that is not
equal to any \$A^r(t)\$, \$t_i\leq
t<t_f$.\\

In this section we have discussed only
binary observables, because they are
simplest and best known. ( One might
take a higher-spin particle in the
Stern-Gerlach case and have more than
two possibilities.) Naturally, owing to
the simplicity of the cases, a usual
Bohrian intuitive discussion is by far
superior in clarity to the expounded
formal one. But it was necessary to
illustrate the concepts in the theory
of sections 3-7. One should appreciate
that this theory covers the general
case.\\

\vspace{0.5cm}

\section{Illustration
for the Extended Entities and Claims}

\noindent Now we discuss a slightly
upgraded version of both the
Mach-Zehnder interference and the
Mach-Zehnder which-way devices (cf
subsection 8.1).

The primary purpose is to illustrate
the {\bf relative character} of the
which-result or interference property.
The secondary purpose is to give an
example for the rest of the entities
and claims that are extended with
respect to Bohr's wave-particle
complementarity.

\vspace{0.5cm}

\subsection{A Slightly
Upgraded Mach-Zehnder Interference
Device}

\begin{figure}

\includegraphics[width=10cm]{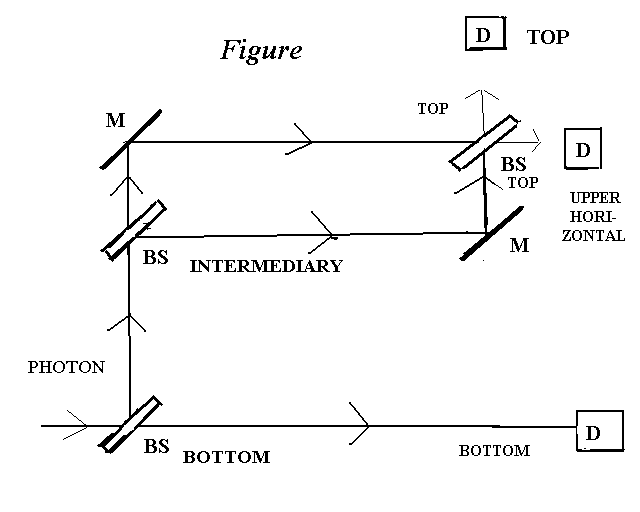}
\caption{In the (slightly upgraded)
Mach-Zehnder interference device there
are three beam splitters (BS): the
bottom one, the intermediary one, and
the top one; there are two mirrors (M),
and three detectors (D): the bottom
one, the upper horizontal one, and the
top (vertical) one. If the top beam
splitter is removed, then we have the
(slightly upgraded) Mach-Zehnder
which-way device. The photon reaches
the bottom beam splitter at the initial
moment \$t_i\$. It is the preparator as
explained in subsection 8.1. For
simplicity we assume that the possible
detection in any of the three detectors
takes place at one and the same moment
\$t_f\$. Let \$t_i\$ be the moment when
the photon passes the intermediary beam
splitter. Naturally, \$t_i<t_0<t_f\$.
There are two observables: Let the
observable \$B\$ be defined at \$t_i\$.
It has two eigen-events: \$Q_h\$,
transmission through the bottom BS and
propagation towards the bottom
detector, and \$Q_v\$ reflection at the
bottom BS and propagation upwards
towards the intermediary BS. The other
observable \$B_0\$ is defined at
\$t_0\$. It has three eigen-events:
\$Q_{lh}\$ lower horizontal propagation
towards the bottom detector, \$Q_{uh}\$
propagation from the intermediary BS
along the upper horizontal line, and
\$Q_v\$, propagation from the
intermediary BS vertically upwards.}
\end{figure}

\vspace{0.5cm}

In spite of the interference in the top
BS, \$B\$ (see the caption of the
Figure) is a 'simplest which-way'
observable and the experiment is a
which-way one relative to \$B\$ in the
sense of Section 5. Namely, due to the
interference, there are only two
detections (with probability one half
each): in the bottom and in the top
detector. If we manipulate the bottom
BS as a preparator (cf subsection 8.1),
then one can easily see that the
necessary and sufficient condition in
Theorem 1 is satisfied. On account of
the simplicity of the experiment, one
can argue also without Theorem 1 as
follows.

Let the localization events be \$P_b\$
in the bottom detector and \$P_t\$ in
the top one respectively. Since
\$U(t_f-t_i)^{\dag}P_b
U(t_f-t_i)=Q_h\$, and
\$U(t_f-t_i)^{\dag}P_t U(t_f-t_i)=
Q_v\$ (the Mach-Zehnder interference
device is time-reversal symmetric), we
can attach equal eigenvalues (which are
anyway not important here) to \$A\$ and
\$B\$, and obtain \$B=A^r(t_i)\$. Thus,
we are dealing with a 'simplest
which-way' experiment in relation to
\$B\$. But ut is not particle-like
behavior in the sense of Bohr because
the photon is not localized all along
\$t_i<t<t_f\$; it exhibits wave-like
behavior in the interval \$t_0\leq
t<t_f$.

This discussion gives rigorous
justification to the intuitive
inference from the Figure that if the
photon ends up in the bottom (top)
detector, it had to come from its
transmission through (reflection at)
the bottom BS.\\

The experiment is an interference one
with respect to the observable \$B_0\$
(see the Caption). This is so because
of the interference in the top BS.\\

Let us define the function (coarsening)
\$B'_0\$ of \$B_0\$ that keeps
\$Q_{lh}\$ as one of its eigen-events
and has \$\Big(Q_{uh}+Q_v\Big)\$ as the
(only) other. The experiment is a
which-way one (again in spite of the
interference in the top BS) in relation
to \$B'_0\$. It is not called
'simplest' because this observable is
defined at \$t_0\$, and
not at \$t_i$.\\

\vspace{0.5cm}

\subsection{The Slightly
Upgraded Mach-Zehnder Which-Way Device}

\noindent Now the observable \$B\$ is
not the 'simplest which-way' one.
Namely, the condition in Theorem 1 is
not satisfied as seen from the fact
that \$Q_v\$ does not lead with
certainty to one detector localization.

Incidentally, the Figure and classical
intuition would suggest that both the
state of reaching the top detector and
that of reaching the upper horizontal
one 'come from' the photon state
\$\ket{\phi_v}\$ corresponding to
\$Q_v\$ (\$Q_v\ket{\phi_v}=\ket{\phi_v}
\$). But this is a false conclusion
because 'come from' should mean
quantum-mechanically the application of
\$U(t_f-t)\$. Or equivalently,
\$\ket{\phi_v}\$ should be obtained
from the two mentioned final states by
application of \$U(t_f-t)^{\dag}\$. But
a unitary operator cannot map
orthogonal states into one and the same
one.

On the other hand, it is seen that
\$U(t_f-t_i)Q_vU(t_f-t_i)^{\dag}=
P_t+P_{uh}\$, and \$P_{uh}\$ is the
event of localization in the upper
horizontal detector. Further,
\$U(t_f-t_i)Q_hU(t_f-t_i)^{\dag}=P_b\$.
Thus, if we take the function (or
coarsening) \$f(A)\$ of the measured
observable \$A\$ defined by the
right-hand-sides, which is
simultaneously also measured in the \M
of \$A\$, then \$B^f\equiv
U(t_f-t_i)BU(t_f-t_i)^{\dag}=f(A)\$.
Since obviously \$[f(A),A]=0\$, the
condition in Theorem 2 is satisfied.
Therefore, \$B\$ is a 'simplest
which-way' observable for \$f(A)$.\\

As to \$B_0\$, the condition in Theorem
1 makes it obvious that it is a
'simplest which-way' observable for
\$A\$. But again it is not so in the
sense of Bohr, because \$B_0\$ is
defined at the moment \$t_0\$, and not
at the initial moment.\\

\vspace{0.5cm}

\section{What is Really Happening?}

\noindent In an attempt to comprehend
what was 'real' for Bohr, let us read
another quotation from him \cite{B5}
\begin{quote}
{\bf Quote BOHR6:} "As a more
appropriate way of expression I
advocate the application of the word
{\it phenomenon} exclusively to refer
to the observations obtained under
specified circumstances, including an
account of the whole experimental
arrangement.(The italics are Bohr's.)
\end{quote}

I think that by "phenomenon" Bohr means
a {\it real phenomenon}, i. e., that
this is where 'reality' enters the
scene in the view of Bohr.

The retrospective observable \$A^r(t)\$
in the visualization theory of Sections
3-7 is primarily a mathematical
construction, an 'evolving' backward in
time of the real observable \$A\$. Even
when the condition in Theorem 1 is
satisfied, and we have the possibility
of a visualization in a 'simplest
which-result' experiment, the
eigen-events \$P^r_n\$ {\it do not
really occur}, not even in the Bohrian
sense. They are {\it only imagined} or
visualized to create a quasi-classical
picture about what is going on {\it
within the experiment} on hand.

This is particularly clear in Wheeler's
delayed-choice experiments, in which
(both in the two-slit
\cite{Wheeler2slits} and in the
Mach-Zehnder \cite{WheelerMZ} cases)
the choice whether the experiment is
going to be the 'which-way' one or the
'interference' one is made {\it after}
the photon has passed the two slits (or
the first beam splitter). Thus, whether
the photon is going one way or both
ways appears to be decided backwards in
time. Obviously, these are not real
events happening in nature.\\

Turning to Bohr's {\it forbiddance} to
{\it combine visualizations from
complementary experiments}, contained
in the complementarity principle, it
seems a justified warning that should
save us from taking the visualizations
too seriously, i. e., as if they were
real occurrences in nature, and thus
different aspects of reality that
should be combined into a complete
picture. In \QMl insight all aspects
are present simultaneously. If one
overemphasizes and even falsifies two
distinct aspects (the wave-like and the
particle-like one in Bohr's approach),
it is natural that
they become incompatible.\\

\vspace{0.5cm}

\section{Assessment}

\noindent A good deal of physical
evaluation of wave-particle \CY in the
Bohrian way was accomplished during the
critical derivation in sections 3-7
because it was done purposely pointing
out the arbitrary or imaginary steps.

Now I'll pay additional attention only
to the (most important) case of
'simplest interference-which result'
experiment, in particular when
\$A^r(t_i)\$ is physically meaningful.

Let me discuss {\bf the first illusion}
(the first drastic imagined step of
changes, cf Section 5, second passage).
As far as the experiment is concerned,
the initial state \$\rho(t_i)\$ can be
replaced by {\it the butchered mixture}
with respect to the 'back-evolved'
observable \$A^r(t_i)\$ (cf (8a)). But,
if there is {\bf coherence} (cf
Definition 1), then the 'which-result'
visualization in case the individual
experiment gives one of these results,
grossly violates the coherence, which,
from the point of view of the
'reality-of-state' approach, to which
this author adheres, is a serious {\bf
falsification} of reality.

Thus, in case of the Mach-Zehnder
which-way device (cf Subsection 8.1)
with a coherent initial state (cf
(13a)), the Bohrian particle-like
aspect creates the {\it illusion} that
the photon takes one of the paths. This
violates grossly the coherence.

John Bell had apparently strong
feelings about this as it is clear from
his term "butchered" state vector for
\$\rho_M(t_i)\$ \cite{Bell1}.

Let me turn to {\bf the second
illusion} (the second drastic imagined
step of changes). Classically,
interpretation of \$\rho(t_i)_M\$ as a
simple mixture would mean that there
is, e. g., a subensemble
\$\rho(t_i)_M^n\$ in it which has the
sharp value in question, and the
individual quantum system belongs to
this subensemble. Then, at first
glance, the 'which-result
visualization' might appear to
correspond to reality.

This argument may stem from a Bohrian
devotedness to classical physics. From
the point of view of the
'reality-of-state' approach, this is an
unacceptable {\it prejudice}. Namely,
as well known, even if the density
operator has no more than a
two-dimensional range, there are
infinitely many decompositions into
density operators, i. e., it can be
written in that many ways as a mixture.
Quantum-mechanically none of them has a
privileged role, which would enable one
interpret it as the {\it real} state of
affairs (as far as decomposition of
ensemble into individual-system states
is concerned).

Neither this point has escaped Bell's
attention as seen from his words
\cite{Bell2}:

\begin{quote}
{\bf Quote BELL2:} "The idea that
elimination of coherence, in one way or
another, implies the replacement of
'and' by 'or' is a very common one
among solvers of the 'measurement
problem'. It has always puzzled me."
\end{quote}

Thus, the Bohrian approach in terms of
complementary particle-like and
wave-like experiments {\bf does not
reveal} two, mutually exclusive, {\bf
aspects} of the state \$\rho(t_i)\$. It
gives a completely distorted view. {\bf
It has very little to do with reality.}
(Though, it does give a simplified
semi-classical understanding of what is
going on in the experiment in partial
agreement with \qm .)\\

As to merits of the present study (if
any), I would like to quote Fagundes
\cite{Fagundes}:
\begin{quote}
{\bf Quote FAGUNDES:} "... physics
progresses by increasing degrees of
abstraction. This is only natural since
'concrete' ideas are just those of our
too limited ordinary sense experience."
\end{quote}

I think, these words are applicable, to
some extent, also to the slight
progress achieved by deriving a sharp
and extended form of
'interference-which-result' \CY from
\QM in this article.\\

\vspace{0.5cm}

\section{Concluding Remarks}

\noindent The derivation in sections
3-7 follows Bohr's endeavor to envelop
the understanding of a quantum
experiment in classical physics as in a
chocolate coating. This is not
surprising when one takes into account
what a low opinion Bohr had of the \Q
formalism. I'll give two excerpts to
illustrate this claim. The first is
from Saunders
\cite{Saunders}.

\begin{quote}
{\bf Quote SAUNDERS:} "The quantum
formalism is {\it only} an abstract
calculus. As we have seen, Bohr made
this point over and over again."
(Italics by Saunders)
\end{quote}

On Bohr's suspicion about the quantum
formalism we have similar words by
Heisenberg \cite{Heis}
\begin{quote}
{\bf Quote HEISENBERG:} "I noticed that
mathematical clarity had in itself no
virtue for Bohr. He feared that the
formal mathematical structure would
obscure the physical core of the
problem,..."\\
\end{quote}

Reichenbach \cite{Reich} made a
variation on Bohr's idea of
visualization by introducing
'interphenomena'.

Fagundes \cite{Fagundes}, probably
laboring under the burden of lack of
sufficient reality in both Bohr's and
Reichenbach's concepts, suggested to
replace visualization by literally
nothing. (A consistently positivistic
point of view, so it appears.)

Holladay's "which-value-interference
\cy" approach \cite{Hollad} is closest
to mine. (I was even influenced by his
terminology.)

I am certain that there are other
praiseworthy related endeavors that
have escaped my attention.\\

Murdoch, in his detailed study of Bohr
\cite{Murdoch}, writes (beginning of p.
68):

\begin{quote}
{\bf Quote MURDOCH1:} "Bohr came to
hold that the wave and particle models
are equally necessary for a complete
description of the real nature of
micro-physical entities - the symmetry
thesis, as I shall call it."
\end{quote}

Later on (in the second passage of p.
79) he writes:
\begin{quote}
{\bf Quote MURDOCH2:} "The symmetry
thesis, then, is difficult to sustain,
and with it the thesis of wave-particle
\cy . The thesis has lost the {\it
palliative value} it once had, and has
now merely a historical significance."
(Emphasis by F. H..)
\end{quote}

The present study confirms this
dismissal of Bohr's \CY \PR on part of
Murdoch (who, as it seems, has studied
Bohr thoroughly). {\bf Present-day
research on the foundations of quantum
mechanics does not need palliation. Its
aim is to understand quantum
reality as it is.}\\

Finally, I would like to point out that
Bohr and the Copenhagen interpretation
\cite{Stapp} {\it caused a substantial
delay in the historical development of
the foundations of \qm .} Bohr's own
words \cite{B6} bear witness to this
claim.

\begin{quote}
{\bf Quote BOHR7:} "There is no quantum
world. There is only an abstract
quantum physical description. It is
wrong to think that the task of physics
is to find out how nature is. Physics
concerns what we can say about nature."
\end{quote}

By now it must be obvious to the reader
that the author's attitude in the
analysis in this article is a rebelion
against this view of Bohr. The
reality-of-state approach, to which the
author is partial, stipulates precisely
the opposite: {\bf however abstract, we
must take the quantum-mechanical
description of experiments seriously
because it reveals how nature is. And
no lesser goal is worthy of our
efforts.} We should be able to "say
about nature" how it really is though
only in an approximation that should be
as good as possible.\\

I think that Gell-Mann gave an
impressive criticism of Bohr
\cite{G-M}:

\begin{quote}
{\bf Quote GELL-MANN:} "The fact that
an adequate philosophical presentation
has been so long delayed is no doubt
caused by the fact that Niels Bohr
brainwashed a whole generation of
theorists into thinking that the job
was done fifty years ago"
\end{quote}

Landsman says \cite{Lands} (p. 214)
"Beller \cite{Beller} went further than
any critic before or after her by
portraying Bohr {\it not} as the Gandhi
of 20th century physics (as in Pais,
1991 \cite{Pais}), but rather as its
Stalin, a philosophical dilettante who
knew no mathematics and hardly even
followed the physics of his day, but
who nonetheless managed to stifle all
opposition by a combination of
political manoeuvring, shrewd rhetoric,
and spellbinding both his colleagues
and the general audience by the
allegedly unfathomable depth of his
thoughts (which, according to Beller,
were actually incoherent and
inconsistent)" (italics by F. H.).

Landsman then comments as follows:
"Despite Beller's meticulous  and
passionate arguments, we do not
actually believe Bohr's philosophy of
\QM was such a great muddle after all."
\\

Let me point out, at the end,  that in
spite of the mentioned delay, it seems
to me that {\it Bohr has done mankind
an invaluable service} by saving it
from being hopelessly lost in a
labyrinth searching for objective \QM
at an early stage. Thus, he made
possible the unparalleled swift
development of \QM in atomic,
molecular, solid-state etc. physics, i.
e., a rapid and immense progress of \QM
as a practical science and no less of
\Q technology.\\

\vspace{0.5cm}

{\bf Appendix A: Proof of Theorem 1}

{\bf Proof} {\it Necessity} If the
retrospective observable \$A^r(t)\$ has
a physical meaning, then one can take
\$B(t)\equiv A^r(t),\$ and the
bijection is the identity map. The
required property obviously holds.

{\it Sufficiency} Let
\$B(t)=\sum_kb_kQ_k(t)\$ be an
observable with physical meaning, and
let \$\{\ket{\phi(t)}_{k,l_k}: \forall
k,\forall l_k\}\$ be a complete
orthonormal eigen-basis of \$B(t)\$
satisfying
$$\forall k:\quad
\sum_{l_k}\ket{\phi(t)}_{k,l_k}
\bra{\phi(t)}_{k,l_k}=Q_k(t).\eqno{(A.1)}$$
This makes the vectors
\$\ket{\phi(t)}_{k,l_k}\$ eigen-vectors
of \$B(t)\$ corresponding to the
eigenvalues \$b_k\$, and the index
\$l_k\$ enumerates the multiplicity
(possible degeneracy) of the eigenvalue
\$b_k\$ of \$B\$.

Further, one can define $$\forall k,
l_k: \quad
\ket{\psi(t_f)}_{k,l_k}\equiv
U(t_f-t)\ket{\phi(t)}_{k,l_k}.
\eqno{(A.2)}$$ On account of the
unitarity of \$U(t_f-t)\$, also the
basis
\$\{\ket{\psi(t_f)}_{k,l_k}:\forall
k,\forall l_k\}\$ is orthonormal and
complete.

Relations (A.1) imply that each state
\$\ket{\phi(t)}_{k,l_k}\$ has the
property \$Q_k(t),\$ and, since we
assume validity of the condition in
Theorem 1, the result of the
measurement of \$A\$ in the
corresponding final state
\$\ket{\psi(t_f)}_{k,l_k}\$ is
certainly \$a_{n(k)},\$ i. e.,
$$\forall k,l_k:\quad
\tr(P_{n(k)}\ket{\psi}_{k,l_k}
\bra{\psi}_{k,l_k})=1.$$

We can rewrite this as
$$\forall k,l_k:\quad
\Big(\bra{\psi}_{k,l_k}P_{n(k)}\Big)
\Big(P_{n(k)}\ket{\psi}_{k,l_k}\Big)=
1,$$ implying
$$\forall k,l_k:\quad
\Big(\bra{\psi}_{k,l_k}
P_{n(k)}^{\perp}\Big)
\Big(P_{n(k)}^{\perp}
\ket{\psi}_{k,l_k}\Big)=0,$$ where
\$P_{n(k)}^{\perp}\equiv I- P_{n(k)},\$
\$I\$ being the identity operator.
Further, one obtains
\$P_{n(k)}^{\perp}\ket{\psi}_{k,l_k}=
0\$ (due to positive definiteness of
the norm), and
$$\forall k,l_k:\quad
P_{n(k)}\ket{\psi}_{k,l_k}=
\ket{\psi}_{k,l_k},$$ implying
$$\forall k,l_k:\quad
P_{n(k)}\ket{\psi}_{k,l_k}
\bra{\psi}_{k,l_k}=
\ket{\psi}_{k,l_k}\bra{\psi}_{k,l_k}.$$
Summing out \$l_k\$ for each value of
\$k,\$ and utilizing (A.2) and (A.1),
one obtains
$$\forall k:\quad
P_{n(k)}\Big(U(t_f-t)Q_k(t)
U(t_f-t)^{\dag}\Big)=U(t_f-t)Q_k(t)
U(t_f-t)^{\dag}.\eqno{(A.3)}$$

On the other hand, we have, in view of
(5b) and (A.3),
$$\forall k:\quad
P_{n(k)}=P_{n(k)}I=P_{n(k)}\Big[
\sum_{k'}\Big(
U(t_f-t)Q_{k'}(t)U(t_f-t)^{\dag})\Big)\Big]
=$$
$$P_{n(k)}\Big[\sum_{k'}\Big(P_{n(k')}
U(t_f-t)Q_{k'}(t)U(t_f-t)^{\dag})\Big)
\Big]=
P_{n(k)}\Big(U(t_f-t)Q_k(t)U(t_f-t)^{\dag}
\Big).\eqno{(A.4)}$$ Relations (A.3)
and (A.4) imply
$$\forall k:\quad U(t_f-t)Q_k(t)
U(t_f-t)^{\dag} =P_{n(k)},$$ which is
equivalent to
$$\forall k:\quad Q_k(t)=P_{n(k)}^r(t)$$
(cf (4b)). Hence, \$B(t)=A^r(t).$\hfill $\Box$\\

\vspace{0.5cm}

{\bf Appendix B: Proof of Theorem 2}

First we prove a lemma.\\

\noindent {\bf A.Lemma} The commutation
condition (12) is {\it equivalent} to
$$\forall (k\not= k'),\enskip \forall
n:\quad Q_k^fP_nQ_{k'}^f=0.
\eqno{(B.1)}$$\\

\noindent {\bf Proof} On account of the
well-known fact that two Hermitian
operators with purely discrete spectra
commute if and only if each
eigen-projector of one commutes with
each eigen-projector of the other, (12)
implies (B.1). Conversely, utilizing
the completeness relation (5b), which
is obviously valid {\it mutatis
mutandis} for the spectral
eigen-projectors of \$B^f\$, one can
see that if (B.1) is valid, then
$$\forall k,n:\quad
Q_k^fP_n=Q_k^fP_nI=\sum_{k'}
Q_k^fP_nQ_{k'}^f=Q_k^fP_nQ_k^f.$$
Adjoining this, one obtains
$$\forall k,n:\quad P_nQ_k^f=Q_k^f
P_nQ_k^f.$$ These two relations imply
(12). \hfill
$\Box$\\

\noindent {\bf Proof of Theorem 2} {\it
Sufficiency} of (12) for 'which-result'
behavior. Straightforward calculation
shows that, owing to (12), A.Lemma, and
(B.1),
$$\forall n:\quad\tr\Big(\rho(t_f)P_n\Big)
=\tr\Big(I\rho(t_f)IP_n\Big)=$$
$$\sum_k\sum_{k'}\tr\Big(Q_k^f\rho(t_f)
Q_{k'}^f
P_n\Big)=\sum_k\sum_{k'}\tr\Big(\rho(t_f)
Q_{k'}^fP_nQ_k^f\Big)=$$
$$\sum_k\tr\Big(Q_k^f\rho(t_f)Q_k^f
P_n\Big)=\tr\Big(\rho(t_f)_{M,B}P_n
\Big).$$

{\it Necessity} of (12) for
'which-result' behavior and proof of
claim B). Now we argue that if (12) is
not valid, then \$B\$ is an
'interference' observable. (In this way
we prove both that which-result
behavior implies (12), and claim B).)

Let (12) not be valid. Then \$\exists
k,n,k'\$  such that
\$Q_k^fP_nQ_{k'}^f\not= 0\$ (cf
A.Lemma). Let, further,
$$Q_k^f\equiv\sum_{l_k}\ket{l_k}\bra{l_k},
\eqno{(B.2a)}$$ and
$$Q_{k'}^f\equiv\sum_{l_{k'}}\ket{l_{k'}}
\bra{l_{k'}} \eqno{(B.2b)}$$ be
projector decompositions into ray
projectors (in terms of basis vectors
defined by (B.2a) and (B.2b), though
incompletely in general). Substitution
of (B.2a) and (B.2b) in the inequality
leads to \$\sum_{l_k}\sum_{l_{k'}}
\ket{l_k}\bra{l_k}P_n\ket{l_{k'}}
\bra{l_{k'}}\not= 0.\$ Hence, there
must exist special values \$l_k\$ and
\$l_{k'}\$ such that
$$\bra{l_k}P_n\ket{l_{k'}}\not= 0.
\eqno{(B.3)}$$

Let $$\ket{l_k,t_i}\equiv U(t_f-t_i)
^{\dag}\ket{l_k},\quad \ket{l_{k'},t_i}
\equiv U(t_f-t_i)^{\dag}\ket{l_{k'}}.
\eqno{(B.4)}$$

Finally, let \$\alpha,\enskip \beta\$
be non-zero complex numbers such that
\$|\alpha|^2+|\beta|^2=1\$. We define
the initial state $$\rho(t_i) \equiv
\Big(\alpha\ket{l_k,t_i}+\beta\ket{l_{k'},
t_i}\Big)\Big(\alpha*\bra{l_k,t_i}+\beta*
\bra{l_{k',t_i}} \Big),$$ where the
asterisk denotes complex conjugation.
Then the final state is
$$\rho(t_f)=\Big(\alpha\ket{l_k}+\beta
\ket{l_{k'}}\Big)\Big(\alpha^*\bra{l_k}+
\beta^*\bra{l_{k'}}\Big).$$ As to the
'unbutchered' and the 'butchered'
states, (B.2a) and (B.2b) imply, as
easily seen,
$$\tr\Big(
\rho(t_f)P_n\Big)=\tr\Big(I\rho(t_f)IP_n
\Big)=$$ $$\sum_{k''}
\tr\Big(Q_{k''}^f\rho(t_f)Q_{k''}^fP_n
\Big)+\sum_{k''\not=
k'''}\tr\Big(Q_{k''}^f
\rho(t_f)Q_{k'''}^fP_n\Big)=$$ $$
\tr\Big(\rho(t_f)_{M,B}P_n\Big)+
\alpha\beta*\bra{l_{k'}}P_n\ket{l_k}+
\beta
\alpha*\bra{l_k}P_n\ket{l_{k'}}.$$ It
follows from (B.3) that this is
different than
\$\tr\Big(\rho(t_f)_{M,B}P_n\Big),\$ i.
e., the experiment distinguishes the
the 'unbutchered' and the 'butchered'
states. Since \$B(t)\$ and \$B^f\$ are
by assumption physically meaningful, so
are, in principle, also the
eigen-states \$\ket{l_k}\$ and
\$\ket{l_{k'}}\$ of \$B^f\$. \hfill $\Box$\\


\vspace{0.5cm}




\begin{thebibliography}{99}

\bibitem{FHinterf}
Herbut F., Vuji\v{c}i\'{c} M.:
First-quantisation quantum-mechanical
insight into the Hong-Ou-Mandel
two-photon interferometer with
polarizers and its role as a quantum
eraser. Phys. Rev. A {\bf 56}, 1-5
(1997)

\bibitem{FHScully1}
Herbut, F.: On EPR-Type Entanglement in
the Experiments of Scully et al. I. The
Micromaser Case and Delayed-Choice
Quantum Erasure. Found. Phys. {\bf 38},
1046-1064 (2008); arXiv:0808.3176

\bibitem{FHScully2}
Herbut, F.: On EPR-Type Entanglement in
the Experiments of Scully et al. II.
Insight in the Real Random
Delayed-Choice Erasure Experiment.
arXiv:0808.3177

\bibitem{Einstein}
Einstein, A.: Remarks on the Essays
Appearing in this Collective Volume.
In: Schilpp, P. A. (ed.) Albert
Einstein: Philosopher-Scientist. The
Library of Living Philosophers Inc.,
Evanston, Illinois (1949), p. 674, last
passage

\bibitem{Born}
Born, M.: The Restless Universe. Dover,
New York (1951), p.283

\bibitem{Home}
Ghose, P., Home, D.: The Two-Prism
Experiment and Wave-Particle Duality of
Light. Found. Phys. {\bf 26}, 943-953
(1996)

\bibitem{other1}
Rangwala, S., Roy, S.M.: Wave Behavior
and Noncomplementary Particle Behavior
in the Same Experiment. Phys. Lett. A
{\bf 190}, 1-4 (1994)

\bibitem{other2}
Ghose, P., Sinha Roy, M.N.: Confronting
the Complementarity Principle in an
Interference Experiment. Phy. Lett. A
{\bf 161}, 5-8 (1991)

\bibitem{B1}
Bohr, N.: Discussion with Einstein on
Epistemological Problems in Atomic
Physics. In: Schilpp, P. A. (ed.)
Albert Einstein: Philosopher-Scientist.
The Library of Living Philosophers
Inc., Evanston, Illinois (1949). P.
222, the passage in the middle of the
page

\bibitem{MZ}
Nachman, P.: Mach-Zehnder
Interferometer as an Instructional
Tool. Am. J. Phys. {\bf 63}, 39-43
(1995)

\bibitem{B2}
Bohr, N.: Natural Philosophy and Human
Cultures. In: Atomic Physics and Human
Knowledge. Wiley, New York (1958), p.
26

\bibitem{B3}
Bohr, N.: Discussion with Einstein on
Epistemological Problems in Atomic
Physics". In: Schilpp, P. A. (ed.)
Albert Einstein: Philosopher-Scientist.
The Library of Living Philosophers
Inc., Evanston, Illinois (1949), p.
210, end of first passage

\bibitem{WheelerMZ}
Wheeler, J. A.: The 'past' and the
delayed-choice double-slit experiment.
In: Marlow A. R. (ed). Mathematical
Foundations of Quantum Theory. Academic
Press, New York (1978), p. 32

\bibitem{B4}
Bohr, N.: Nature (Suppl.) {\bf 128},
691 (1931)

\bibitem{Schlossh}
Schlosshauer, M.: arXiv: 0804.1609

\bibitem{Reich}
Reichenbach, H.: Philosophic
Foundations of Quantum Mechanics. Univ.
California Press, Berkeley (1965)


\bibitem{FHcoh}
Herbut, F.: A Quantum Measure of
Coherence and Incompatibility. J. Phys.
A: Math. Gen. {\bf 38}, 2959-2974
(2005)


\bibitem{Bell1}
Bell, J.: Against 'Measurement'. Phys.
World, August, p. 37, right column,
second passage


\bibitem{BLM}
Busch, P., Lahti, P.J., Mittelstaedt,
P.: The Quantum Theory of Measurement.
Lecture Notes in Physics, M2. Springer
Verlag, Berlin (1991).


\bibitem{FHmeas}
Herbut, F.: Delayed-Choice Experiments
and Retroactive Apparent Occurrence in
the Quantum Theory of Measurement.
Found. Phys. {\bf 24}, 117-137 (1994),
section 2

\bibitem{FHback}
Herbut, F.: On Retroactive Occurrence
and Twin Events in Quantum Mechanics.
Found. Phys. Lett. {\bf 9}, 437-446
(1996)


\bibitem{Solvey}
Bacciagaluppi, G., Valentini, A.:
Quantum Theory at the Crossroads.
Cambridge Univ. Press, Cambridge
(2006). arXiv:quant-ph/0609184

\bibitem{origScully2}
Kim Y.-H., Yu R., Kulik S. P., Shih Y.,
Scully M. O.: Delayed "choice" quantum
eraser. Phys. Rev. Lett. {\bf 84}, 1-5
(2000); also arXiv: quant-ph/9903047


\bibitem{realMZ}
Grangier, P., Roger, G., Aspect, A.:
Experimental Evidence for a Photon
Anticorrelation Effect on a Beam
Splitter: A New Light on Single-Photon
Interferences. Europhys. Lett. {\bf 1},
173-179 (1986)


\bibitem{Wheeler2slits}
Wheeler J. A.: The 'past' and the
delayed-choice double-slit experiment.
In: Marlow A. R. ed. Mathematical
Foundations of Quantum Theory. Academic
Press, New York (1978), p. 11


\bibitem{B5}
Bohr, N.: Discussion with Einstein on
Epistemological Problems in Atomic
Physics. In: Schilpp, P. A. (ed.)
Albert Einstein: Philosopher-Scientist.
The Library of Living Philosophers
Inc., Evanston, Illinois (1949). Pp.
237-238

\bibitem{Bell2}
Bell, J.: Against 'Measurement'. Phys.
World, August, p. 36, right column,
fourth passage from below

\bibitem{Fagundes}
Fagundes, H.V.: The Discontinuous
Spatiality of Quantum Mechanical
Objects. arXiv:quant-ph/0504125v2

\bibitem{Saunders}
Saunders, S.: Complementarity and
Scientific Rationality. Found. Phys.
{\bf 35}, 417-447 (2005), p. 432

\bibitem{Heis}
Rosenthal, S. (ed.): Niels Bohr, His
Life and Work as Seen by his Friends
and Colleagues. New York (1967), p. 98

\bibitem{Hollad}
Holladay, W.G.: The Nature of
Particle-Wave Complementarity. Am. J.
Phys. {\bf 66}, 27-33 (1998)

\bibitem{Murdoch}
Murdoch, D.: Niels Bohr's Philosophy of
Physics. Cambridge Univ. Press,
Cambridge (1987)

\bibitem{Stapp}
Stapp, H.P.: The Copenhagen
Interpretation. Am. J. Phys. {\bf 40},
1098-1116 (1972)

\bibitem{B6}
Petersen, A.: The Philosophy of Niels
Bohr. The Bulletin of the Atomic
Scientists. September (1963), p. 8

\bibitem{G-M}
Gell-Mann, M.: The Nature of the
Physical World. Wiley, New York (1979)

\bibitem{Lands}
Landsman, N.P.: When Champions Meet:
Rethinking the Bohr-Einstein Debate.
Stud. Hist. Phil. Mod. Phys. {\bf 37},
212-242 (2006), pp. 214-215

\bibitem{Beller}
Beller, M.: Quantum Dialogue. Univ.
Chicago Press, Chicago (1999)

\bibitem{Pais}
Pais, A.: Niels Bohr's Times: In
Physics, Philosophy, and Polity. Oxford
Univ. Press, Oxford (1991)


\end{thebibliography}
\end{document}